\documentclass[aps,amsmath,amssymb,floatfix,showpacs]{revtex4}
\usepackage{graphicx,color, bm} 
\usepackage{dcolumn} 

\def\ADD#1{{\textcolor{blue}{#1}}}         

\begin{document}
\title 
{Isotropisation at small scales of rotating helically-driven turbulence}

\author{P.D. Mininni$^{1,2}$, D. Rosenberg$^1$ and A. Pouquet$^1$}
\affiliation{
$^1$ National Center for Atmospheric Research, PO Box 3000, Boulder, CO 80307 USA\\
$^2$ Departamento de F\'\i sica, Facultad de Ciencias Exactas y Naturales and IFIBA, CONICET, Ciudad Universitaria, 1428 Buenos Aires, Argentina.
}

\begin{abstract}
We present numerical evidence of how three-dimensionalization occurs at small scale in rotating turbulence with Beltrami (ABC) forcing, creating helical flow. The Zeman scale $\ell_{\Omega}$ at which the inertial and eddy turn-over times are equal is more than one order of magnitude larger than the dissipation scale, with the relevant domains (large-scale inverse cascade of energy, dual regime in the direct cascade of energy $E$ and helicity $H$, and dissipation) each moderately resolved. These results stem from the analysis of a large direct numerical simulation on a grid of $3072^3$ points, with Rossby and Reynolds numbers respectively equal to $0.07$ and $2.7\times 10^4$. At scales smaller than the forcing, a helical wave-modulated inertial law for the energy and helicity spectra is followed beyond $\ell_{\Omega}$ by Kolmogorov spectra for $E$ and $H$. Looking at the two-dimensional slow manifold, we also show that the helicity spectrum breaks down at $\ell_{\Omega}$, a clear sign of recovery of three-dimensionality in the small scales.

\ADD{{\it To appear, Journal of Fluid Mechanics, 2012}}
\end{abstract}
\maketitle


\section{Introduction}\label{sec:intro}

Many studies have been devoted to rotating flows because of their prevalence in geophysical and astrophysical settings, and also because they present one clear example of how coherent structures, turbulent eddies and waves interact, and of how turbulence statistics is affected by such interactions. For example, it was shown in \cite{hopfinger82} (see also \citealt{hopfinger93} for a review) that, as expected by the Taylor-Proudman theorem \citep{proudman, taylor} and its recent dynamical extension \citep{waleffe}, vortical structures organize in the direction of rotation. Strictly speaking, the Taylor-Proudman theorem shows that columns, or rather two-dimensional (2D) flows, are steady-state solutions of the equations in the presence of rotation. However, this result does not explain how the system evolves dynamically towards these states; the argument in \cite{waleffe} extends the result to take into account the time evolution.

Recent laboratory experiments have been able to study such flows in more detail. For example,  the development of anisotropy and cyclonic and anti-cyclonic vortices is studied in  \cite{moisy11} for freely decaying rotating flows. The statistical properties of rotating flows can vary significantly, as seen when examining the scaling of high-order structure functions of velocity gradients, shown in \cite{simand}, \cite{baroud03}, \cite{moisy}, and \cite{van}. Although the origin of such differences in scaling is not clear, studies using direct numerical simulations (DNS) suggest that the presence of helicity may be one of the causes of these variations, as may be the strength of rotation (as measured by the Rossby number), and the range of scales in which measurements are made (direct cascade to small scales versus inverse cascade to large scales).

Helicity (i.e., velocity-vorticity correlations) is an invariant of the ideal equations of motion  (see \citealt{moffatt_tsinober} for a review). The dynamical effects of helicity in rotating flows have not been the topic of intense studies until recently. However, it was postulated early on that helicity may play a role in the stabilization of rotating convective storms in the atmosphere \citep{lilly}, and it has been measured recently in the atmosphere \citep{koprov}, with a spectral  law close to that of Kolmogorov, both during the day and night. The influence of helicity on spectral laws was studied in \cite{chakra07}, whereas \cite{orlandi} showed that in a helical rotating pipe flow, drag reduction occurs through the modification of near-wall structures with spiral motions entraining streamwise vorticity. Although in a maximally helical flow the Lamb vector {$\mbox{\boldmath $\omega$} \times {\bf u}$ is equal to zero (where $\mbox{\boldmath $\omega$}$ is the vorticity, and {\bf u} the velocity)}, in turbulent helical flows nonlinear interactions are drastically reduced but not cancelled. This is because, as was shown early on by \cite{rhk} using a derivation of statistical equilibrium ensembles, the energy cascade cannot take place with maximal helicity at all scales, and therefore relative helicity 
\begin{equation}
r(k) = \frac{H(k)}{kE(k)}
\end{equation}
(where H and E are respectively the helicity and the energy) must decrease with wavenumber. Moreover, it is known that Beltrami flows are unstable \citep[see, e.g.,][]{podvigina}.

In spite of this reduction of nonlinear interactions, in the absence of rotation the presence of helicity only results in a delay in the onset of energy decay: the maximum of enstrophy (integral of the square vorticity) occurs at later time, but otherwise the energy decay proceeds at the same rate as in the non-helical case. This is to be expected as it takes a longer time for the cascade to begin, but once it is established the cascade rate is the same (see \citealt{chen} and references therein for numerical studies of helical turbulence in the absence of rotation). On the other hand, the decay rate is greatly reduced in the presence of both rotation and helicity \citep{teitelbaum}.

In this paper we investigate numerically the co-existence of several dynamical regimes in forced rotating turbulence and, specifically, we focus on the transition between a large-scale anisotropic inertial range dominated by rotation, which could be called the ``Coriolis subrange,'' and an isotropic inertial subrange (the ``Kolmogorov subrange'') associated with the return to classical non-rotating Kolmogorov nonlinear interactions, as hypothesized for example in \cite{zeman} and \cite{canuto}. {Note that although we will consider here the particular case of helical rotating turbulence, in the following we use the terms ``Coriolis subrange'' and ``Kolmogorov subrange'' independently of whether the turbulence has net helicity or not.}

A thorough discussion of the departure from isotropy in rotating flows can be found in \cite{cambon_89} (with recent reviews in \citealt{cambon_rev, cambon_04}, and \citealt{bellet_06}). However, the problem of recovery of isotropy at small scales (in both helical and non-helical flows) is seldom studied, due in part to the experimental difficulty in quantifying the properties of small-scale vortices, and due to the numerical one of resolving several inertial subranges, both to small and large scales. Phenomena associated with large scales, as the inverse cascade of energy in strongly rotating flows  \citep{smith96, smith99, chen05}, or the generation of zonal flows by Rossby wave resonances \citep{newell}, have been studied extensively, whereas the interplay between initially two-dimensional structures becoming three-dimensional (3D) and vice-versa was studied in \cite{bartello_CS}, by varying the rotation rate in numerical simulations with hyper-viscosity. Note also that \cite{naza} have claimed, using phenomenology and critical balance arguments between the inertial time and the eddy turn-over time, that a recovery of isotropy should take place at sufficiently small scales. {However, \citet{Lamriben} recently were able to measure anisotropic energy transfers in an experiment of freely decaying rotating turbulence and found anisotropy to be more pronounced at smaller scales.}

Here, the recovery of isotropy at small scales {in forced rotating turbulence} is studied using a DNS as described in \S \ref{s:dns}. A simulation at unprecedented resolution allows us to resolve and identify both ranges, and the transition from one to the other. The phenomenological context for non-helical rotating flows, developed in \cite{dubrulle}, \cite{zeman}, \cite{zhou}, and \cite{canuto}, is  extended to the helical case in \S \ref{s:aniso2}; numerical results are described in \S \ref{s:res}, while in \S \ref{s:conclu} brief conclusions are presented.

\section{Numerical set-up}\label{s:dns}

The simplest way to include rotation in a turbulent flow, and to prevent effects of an Ekman layer from affecting the overall flow, is to consider solid body rotation with planar geometry using periodic boundary conditions. We thus have performed a DNS run of rotating helical turbulence on a grid of $N^3=3072^3$ points using $\approx 2\times 10^4$ processors, with a box size $L_0=2\pi$ (corresponding to a wavenumber $k_{min}=1$), and with forcing at $k_F/k_{min}=4$. Even though such a choice of forcing wavenumber precludes the development of a sizable inverse cascade, it nevertheless allows for sufficient energy transfer to the large scales, rendering the energy transfer to small scales presumably small (see below). 

Detailed information about the specific procedure and the code can be found in our previous papers dealing with studies of rotating turbulence performed at lower resolution, and in which isotropy was not allowed to develop fully at small scale \citep{rot512,1536a,1536b,phil_trans}. These papers discuss phenomenological theories for rotating turbulence with and without helicity \citep{phil_trans}, results from numerical simulations showing evidence of a distinct scaling of the energy and the helicity in rotating helical flows \citep{rot512, 1536a}, and evidence of the absence of intermittency in such flows \citep{1536b}. In the present paper, on the other hand, the focus is on the recovery of isotropy at small scales when the Reynolds number is large enough. For the purpose of completeness, we present again in this manuscript the scaling of energy and helicity spectra in the Coriolis subrange obtained in the simulations. We also present new phenomenological results associated with a requirement that the Zeman wavenumber for both the energy and helicity spectra be the same, and numerical evidence confirming our prediction. This is to our knowledge the first time that the isotropization of the small scales in a direct numerical simulation of forced rotating turbulence is achieved. Also note that the phenomenological understanding of helical rotating turbulence is confirmed by this large computation.

The equations to be integrated numerically using a pseudo-spectral code are written in the rotating frame for the velocity field and the vorticity $\mbox{\boldmath $\omega$} = \nabla \times {\bf u}$:
\begin{equation}
\frac{\partial {\bf u}}{\partial t} + \mbox{\boldmath $\omega$} 
    \times {\bf u} + 2 \mbox{\boldmath $\Omega$} \times {\bf u} =
    - \nabla {\cal P} + \nu \nabla^2 {\bf u} + {\bf F_{\rm ABC} } \ ,
\label{eq:momentum}
\end{equation}
where ${\cal P}$ is the total pressure  modified by the centrifugal term, and is obtained self-consistently by taking the divergence of Eq.~(\ref{eq:momentum}), assuming incompressibilty, $\nabla \cdot {\bf u}=0$. The rotation $\mbox{\boldmath $\Omega$}$ is imposed in the vertical direction. The Reynolds number at the onset of the inverse cascade is $\Re=U_0L_F/\nu\approx 2.7\times 10^4$ (with $U_0$ the r.m.s.~velocity, $L_F=2\pi/k_F$ the forcing scale, and $\nu=6.5 \times 10^{-5}$ the kinematic viscosity), and the Rossby number is $Ro= U_0/[L_F \Omega] \approx 0.07$. The micro Rossby number is defined as the ratio of r.m.s.~vorticity to imposed rotation, $Ro_{\omega}=\omega_{rms}/\Omega$; at the onset of the inverse cascade we find $Ro_{\omega}\approx 4.5$. Such a rather high value can be associated with the isotropization of small-scales, with the enstrophy spectrum peaking at the dissipation wavenumber since one recovers at small scales a Kolmogorov law (see \S \ref{s:res}). When computed using the r.m.s.~vorticity at the Zeman scale (see \S \ref{s:aniso2}), the micro-Rossby number is found to be of order unity. As explained in \citet{cambon97}, the value of the micro-Rossby number plays a central role in the development of anisotropies and in the determination of the velocity derivative skewness; in practice, $Ro_{\omega}$ should be unity or larger for nonlinear interactions not to be completely damped by scrambling effects of inertial waves. Finally, the Reynolds number computed at the Zeman scale is of the order of $400$.

The procedure used to initialize the run is the following. A first simulation is done starting from null initial velocity field (fluid at rest), and with weak rotation. The forcing ${\bf F}_{\rm ABC}$ is a Beltrami (fully helical) ABC acceleration centered on $k_F$:
{\setlength\arraycolsep{2pt}
\begin{eqnarray}
{\bf F}_{\rm ABC} &=& f_0 \left\{ \left[B \cos(k_F y) + 
    C \sin(k_F z) \right] \hat{x} + \right. {} \nonumber \\
&& {} + \left[A \sin(k_F x) + C \cos(k_F z) \right] \hat{y} + 
   {} \nonumber \\
&& {} + \left. \left[A \cos(k_F x) + B \sin(k_F y) \right] 
   \hat{z} \right\},
\label{eq:ABC}
\end{eqnarray}}
with $f_0=0.45$, $A=0.9$, $B=1$, and $C=1.1$ (this choice of the $A$, $B$, and $C$ coefficients allows for a slightly faster instability and development of turbulence, see \citealt{Archontis03}). {Note that in the absence of rotation, the use of ABC forcing results in an approximately isotropic (but not mirror-symmetric) flow, with modes excited in the $k_x$, $k_y$, and $k_z$ axes in Fourier space with approximately equal intensity.} In this first run the imposed rotation $\Omega$ is set equal to $0.06$, and the run is continued for $\approx 3$ turn-over times, in order to let the flow establish a statistically steady state. With $\Omega=0.06$, the resulting value of the Rossby number is so large that the flow is in practice unaware of the imposed rotation, and this choice (as opposed to $\Omega\equiv 0$) allows one to verify that indeed at small rotation rate, the turbulence is unaffected by the rotation. Indeed, at this stage the flow displays an isotropic Kolmogorov inertial range; any anisotropy that develops at large scale once $\Omega$ is increased will thus be related with the effect of rotation, and will happen dynamically.

Once the statistically weak-rotation steady stage is reached, at a time arbitrarily re-labeled $t=0$, $\Omega$ is set equal to $5$. The initial time step is $\Delta t = 2.5 \times 10^{-4}$; it was decreased twice by a factor of $2$, at around $t=2$ and $ t=4$, in order to accommodate the increase in energy due to the inverse cascade. In the following we will mostly focus on the late time state of this run. At late times, the injection rates of energy and helicity are respectively $\epsilon \approx 0.2$ and $\tilde \epsilon \approx 0.8$; thus, $\tilde \epsilon \approx k_F \epsilon$ and indeed helicity injection is close to maximal. On the other hand, the direct cascade fluxes of energy and helicity, $\Pi$ and $\Sigma$, are $\Pi \approx 0.1$ and $\Sigma \approx 0.7$ respectively. The Schwarz inequality $H(k)\le kE(k)$, in terms of energy and helicity spectra $H(k)$ and $E(k)$, also applies to their injection rates: one can force independently the energy (symmetric part of the velocity correlation tensor) and the helicity (anti-symmetric part of the tensor) \citep[see, e.g.,][]{pouquet_patterson}, with the inequality giving a bound on the amount of helicity one can inject. The inequality implies $|r(k)| \le 1$, with maximal helicity when the equality is fulfilled. However, the fluxes themselves are independent of this inequality, as each quantity can {\it a priori} cascade with different cascade times.

The code used for the simulations is parallelized using a hybrid MPI/OpenMP scheme, and applies a $2/3$ de-aliasing rule \citep{ghybrid, Gomez05a}; the temporal scheme is a second-order Runge-Kutta. In the absence of dissipation, the energy $E$ and the helicity $H=\left<{\bf u} \cdot \mbox{\boldmath $\omega$} \right>/2$ are ideal invariants; note that in runs with maximally helical forcing, the isotropic helicity spectrum is of one sign until the dissipation scale. As will be shown later, this property will not hold for other (anisotropic) spectra. Finally, note that no hyper-viscosity is used, and no friction at large scale is introduced either.

\section{The Zeman scale}\label{s:aniso2}

We now recall how the Zeman scale is derived, i.e., the scale at which the eddy turn-over time and the inertial wave time become equal. The phenomenology below is isotropic since the Zeman scale is the scale where isotropy is hypothesized to be recovered; it is thus consistent with the hypothesis to use isotropic phenomenological arguments to derive it. The validity of the hypothesis can be (and is) later confirmed by the large direct numerical simulation analyzed in this paper.

\begin{figure} \begin{center}
\includegraphics[width=13.0cm]{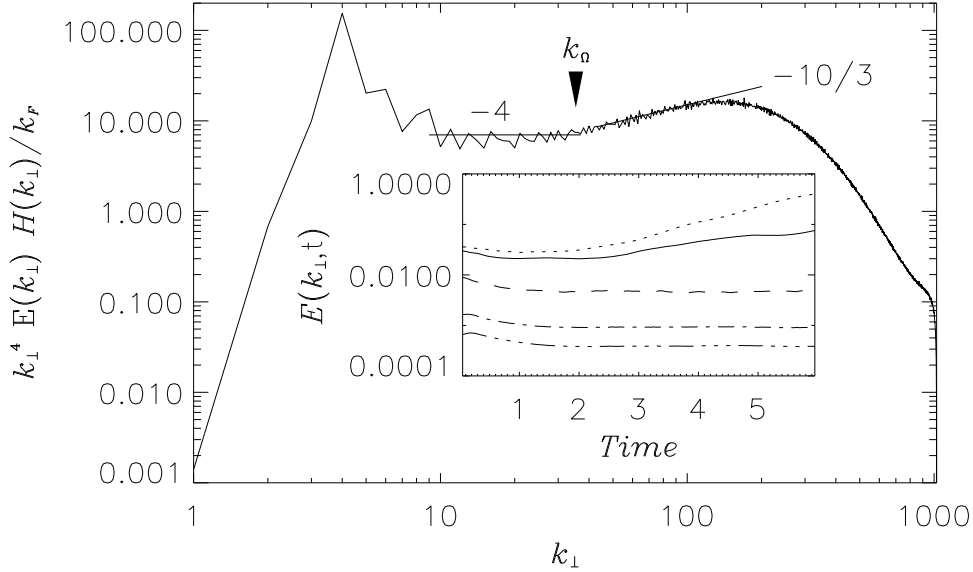}
\caption{Spectral product $E(k_{\perp})H(k_{\perp})k_{\perp}^4/k_F$,  compensated and normalized, as a function of perpendicular wavenumber $k_{\perp}$ and averaged for $5\le t \le 6$;  $k_F=4$ is the forcing wavenumber and $E(k_\perp)$, $H(k_\perp)$ the reduced perpendicular energy and helicity spectra (see text, eq. (\ref{reduced})). Shown for reference are a $-4$ law (horizontal line), as well as a dual Kolmogorov law, $E(k) \sim H(k) \sim k^{-5/3}$ (slanted line). The Zeman wavenumber $k_\Omega = 2\pi/\ell_\Omega$ is indicated by an arrow. The inset shows the temporal evolution of the energy in a few Fourier shells, both in the inverse cascade range ($k_\perp=1$ in solid and $k_\perp=2$ in dotted lines), and in the direct cascade range ($k_\perp=20$ in dashed, $k_\perp=60$ in dash-dotted, and $k_\perp=100$ in dash-triple-dotted lines).}
\label{fig:spec}
\end{center} \end{figure}

One can write the overall energy and helicity spectra as stemming from a combination of a rotation component and of a classical Kolmogorov component. The former, dealing with rotation and helicity, were derived in  \cite{phil_trans} using dimensional analysis (see Table 1, case M09). Omitting constants of order unity, one has:
\begin{equation}
E(k)\sim \epsilon^a \ \tilde \epsilon^{b} \ \Omega^{f}\  k^{-e} + \ \epsilon^{2/3}\ k^{-5/3} \ \ \ , \ \ \ 
H(k)\sim \epsilon^c \ \tilde \epsilon^d \ \Omega^{g}\  k^{-h} +\  \tilde \epsilon \  \epsilon^{-1/3} \ k^{-5/3} \ , 
\label{2_spectra} \end{equation}

The first terms in these expressions correspond to the Coriolis subrange, while the second terms correspond to the Kolmogorov subrange. Assuming constancy with wavenumber of the flux of helicity in the Coriolis subrange, and using that the helicity flux is the ratio of the total helicity divided by a characteristic time, which we associate with the transfer time $\tau_{tr} \sim \tau_{NL}^2/\tau_\Omega \sim \Omega/[k^3 E(k)]$, where $\tau_\Omega$ is the inertial wave time and $\tau_{NL}$ is the eddy turn-over time, we obtain the following relationships:
\begin{equation}
a+c=0, \ b+d=f+g=1, \ b=3-2a-e, \ f=3a+3e-7 \ ,
\label{zeman_G2}\end{equation}
and
\begin{equation}
e+h=4 \ .
\label{zeman_G2b}\end{equation}
The latter two laws in Eq.~(\ref{zeman_G2}) stem from dimensional compatibility in terms of length-scales and time-scales. Equation (\ref{zeman_G2b}) relates the energy and helicity spectral indices, and has been verified in the numerical simulations analyzed in \cite{rot512,1536a}. The actual values of the spectral indices may {\it a priori} depend on the fraction of helicity in the flow; the indeterminacy in equation (\ref{zeman_G2b}) in terms of spectral exponents may be lifted under the assumption of  maximal helicity at all scales, and yields $e=5/2$ and $h=3/2$. Note however that it can be argued that a state of maximal helicity at all scales is not reachable, at least in the framework of statistical mechanics of a truncated system of Fourier modes \citep{rhk}.

It is then straightforward to write an expression for the wavenumber at which there is equality of the inertial wave time and the eddy turn-over time, using for the eddy turn-over time the spectra in Eq.~(\ref{2_spectra}), and with $e<3$ to ensure convergence of energy dissipation:
\begin{equation}
k_{{\Omega}_G}\sim \epsilon^{\alpha}\  \tilde \epsilon^{\beta} \ \Omega^{\delta} \ ,
\label{zeman_G}\end{equation}
with
\begin{equation}
\alpha={\frac{-a}{3-e}}\ \ \  , \ \ \  \beta= {\frac{e+2a-3}{3-e}} \ \ \ , \ \ \delta={\frac{3(3-a-e)}{3-e}} \ .
\label{zeman_G3}\end{equation}

\begin{figure} \begin{center}
\includegraphics[width=13.0cm]{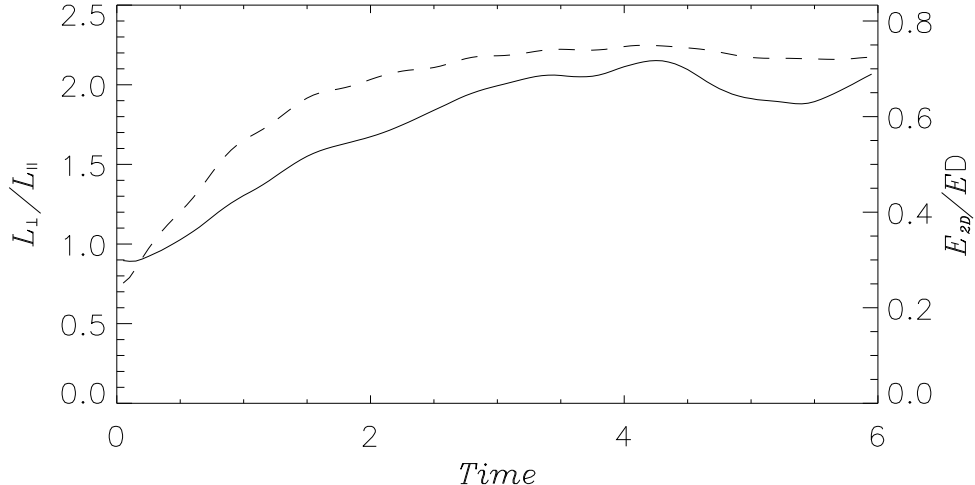} 
\caption{Temporal evolution of the ratio of the perpendicular to the parallel integral scales defined in eq. (\ref{scales}) (solid line, left scale) and of the fraction of the kinetic energy which is in the two-dimensional mode (dashed line, right scale). The initial condition is fully isotropic developed turbulence (see text), and anisotropy grows {subsequently} due to the strong imposed rotation.}
\label{fig:new} \end{center} \end{figure}

Let us call $k_{E}$ and $k_H$ the wavenumbers at which, respectively, the two components of the energy and helicity spectra balance in Eq. (\ref{2_spectra}). Their expressions can be easily found:
\begin{equation}
k_{E}\sim \epsilon^{\kappa}\  \tilde \epsilon^{\lambda} \ \Omega^{\mu}  \  , \ 
k_{H}\sim \epsilon^{\xi}\  \tilde \epsilon^{\rho} \ \Omega^{\psi}  \  , 
\label{zeman_G4}\end{equation}
with
\begin{equation}
\kappa={\frac{3a-2}{3e-5}}\ \ \ , \ \ \ 
\lambda={\frac{3(3-2a-e)}{3e-5}} \  \ \ , \ \ \ 
\mu={\frac{3(3a+3e-7)}{3e-5}} \ ,
\label{zeman_G5}\end{equation}
\begin{equation}
\xi={\frac{3a-1}{3e-7}}\ \ \ , \ \ \ \rho={\frac{3b}{3e-7}} \ \ \ , \ \ \  \psi={\frac{-3(3a+3e-8)}{3e-7}}  \  ,
\label{zeman_G6}\end{equation}
and with $e\not= 5/3$, $e\not= 7/3$.

If we now impose that $k_{{\Omega}_G}=k_E$, this leads to $k_{{\Omega}_G} = \epsilon^{-1/2} \Omega^{3/2}$, with $2a=3-e; \ b=0; \ 2f=3e-5$; then, $k_H=k_{{\Omega}_G}$ follows. Note that the physically plausible condition that $k_{E}$ and $k_{H}$ increase with rotation leads to $e <7/3$. This bound was also derived in \cite{chakra07b} from a different point of view on the basis of a non-rotating helicity cascade following \cite{brissaud}. It is noteworthy that $e<7/3$ excludes a state of maximal helicity for which $e=5/2$.

Plugging these exponents into the expression of the scale at which the inertial wave time and the eddy-turnover time are equal, Eq.~(\ref{zeman_G}), then leads to the same formulation as in the non-helical case as derived by \cite{zeman}, an expression that is independent of the energy spectral index, and hence of the helicity index as well: 
\begin{equation}
k_{\Omega}\sim \epsilon^{-1/2}\Omega^{3/2} \ .
\label{zeman_0} \end{equation}
Under the hypothesis that isotropy is recovered at the same wavenumber for the energy and helicity spectra, we thus arrive at the conclusion that this wavenumber does not depend on the amount of helicity, nor does it depend on the spectral indices $e$ and $h$. We propose here to name this characteristic scale the Zeman scale, so that we can distinguish it from the Ozmidov scale, familiar when studying non-rotating stratified flows and obtained in a similar fashion by using the inverse of the Brunt-Va\"issala frequency instead of the inertial wave frequency. Then, the energy and helicity spectra in the direct Coriolis subrange, with $k_{{\Omega}_G}=k_{\Omega}=k_E=k_H$, simplify into:
\begin{equation}
E(k)\sim \epsilon^{m}\  \Omega^{n}\  k^{-e} \ \ \ , \ \ \ 
H(k)\sim \epsilon^{p}\  \tilde \epsilon \ \Omega^{q} \  k^{e-4} \ , 
\label{2_spectra_simple} \end{equation}
with
\begin{equation}
m={\frac{3-e}{2}}\  \ \ , \ \ n={\frac{3e-5}{2}} \ \ \ , \ \ \ 
p={\frac{-(3-e)}{2}}\  \ \ , \ \ \  q={\frac{7-3e}{2}} \ .
\label{2_spectra_simple2} \end{equation}

It is then straightforward to derive the dependence of the relative helicity spectrum on the energy spectral index:
\begin{equation}
r(k) = \frac{H(k)}{kE(k)} \sim k^{2e-5} \ .
\label{maximal} \end{equation}
The actual prefactor in this expression is given by the relative helicity imposed by the forcing at the forcing scale. In the Coriolis subrange, and as long as $e>2$, the relative helicity thus decreases with wavenumber more slowly than in the isotropic Kolmogorov subrange, where $r(k)\sim 1/k$.

\begin{figure} \begin{center}
\includegraphics[width=13.7cm]{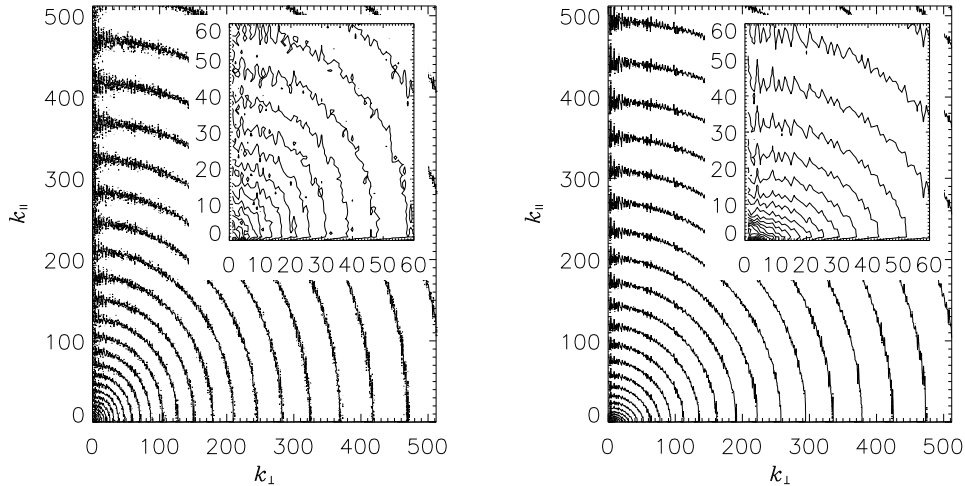}
\caption{{Isocontours of the axisymmetric energy spectrum $e(k_\perp,k_{\parallel})/\sin\theta$ at $t=0$ (left) and after rotation is turned on (right, averaged over $5 \le t \le 6$). Only wavenumbers up to $k_{max}/2 = 512$ are shown (with $k_{max}$ the maximum wavenumber resolved in the computation). The insets in each figure show a detail of the isocontours for small wavenumbers. Near circular contour levels at $t=0$ for all wavenumbers indicate isotropy. At late times, elliptical isocontours are an indication that anisotropy has developed,  {largely} so for small wavenumbers.}}
\label{fig:dev_aniso} \end{center} \end{figure}
  
The final expressions in Eqs.~(\ref{2_spectra_simple}) and (\ref{2_spectra_simple2}) imply $b=0$ in Eq.~(\ref{2_spectra}). One can remark that $b=0$ might possibly have been hypothesized to start with: if $b$ were not equal to zero, $\tilde \epsilon$ being a pseudo-scalar and the energy a scalar, the constant in front of the energy spectrum in Eq. (\ref{2_spectra}) should be a pseudo-scalar as well; it is noteworthy that in fact this condition ($b=0$) arises directly from the analysis under the plausible hypothesis of equality of the Zeman wavenumber and the wavenumber at which the two components of the energy (and helicity) spectra equilibrate. Also, as in the non-rotating case, the cascade of helicity is linear in the rate $\tilde \epsilon$, hence no constraint on the relative energy and helicity fluxes appear from the condition $k_{{\Omega}_G} = k_{\Omega} = k_E = k_H$. As $e$ approaches the value of $7/3$ (respectively, $5/3$), the dependence on rotation becomes linear for the energy spectrum (resp., the helicity spectrum), and is weak for $H(k)$ (resp., $E(k)$).

As stated previously, many studies have been devoted to the case of rotation that is sufficiently strong that anisotropy prevails at all scales since $\ell_{\Omega}\rightarrow 0$ when $\Omega \rightarrow \infty$. This is what is found when using a closure of rotating turbulence as described in detail in \cite{bellet_06} (see, e.g., the general discussion): for the so-called Asymptotic Quasi-Normal Markovian (AQNM) model, developed for strong rotation by removing rapid oscillations, anisotropy increases with wavenumber (see Figs. 5 \& 8, but also note that the Reynolds number of these computations, $\Re\approx 5$, is rather low). Indeed, in numerical simulations, in numerical modeling and/or in laboratory experiments in which the Reynolds number is moderate, it often happens that the Zeman scale is smaller than the dissipation scale and hence anisotropy prevails at all resolved scales in the inertial range. In many of these previous studies anisotropy was reported to increase with wavenumber (see, e.g., \citealt{bellet_06}). {Recent measurements in experiments of freely decaying turbulence \citep{Lamriben} also indicate stronger anisotropy at small scales, with anisotropy prevailing at all scales.}

\begin{figure} \begin{center}
\hskip-0.25truein \includegraphics[width=14.0cm]{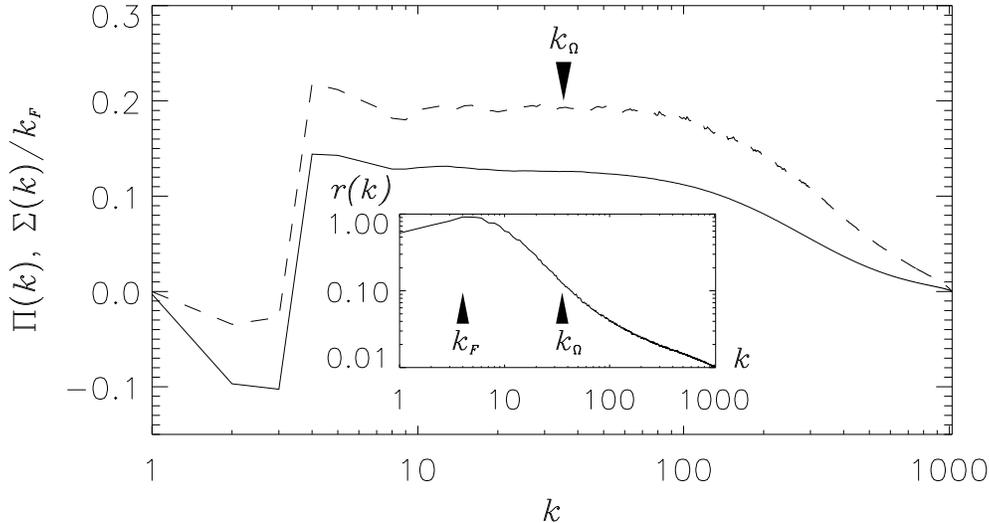}
\caption{Fluxes of energy (solid line) and helicity (dashed line, normalized by the forcing wavenumber, $k_F=4$) as a function of (isotropic) wavenumber, and averaged for $5\le t \le 6$.  The Zeman wavenumber is indicated by an arrow. Note the seamless transition from the anisotropic to the isotropic regime at small scale. Similar results obtain when plotting the fluxes against $k_{\perp}$. In the inset is given the spectrum of relative helicity $r(k)=H(k)/[kE(k)]$, with the forcing and Zeman wavenumbers, $k_F$ and $k_{\Omega}$,  indicated by arrows.}
\label{fig:flux} \end{center} \end{figure}

By contrast, the present work is devoted to the issue of resolving the Zeman scale and below, so that the isotropic inertial range can be unraveled; this implies high Reynolds number computations for realistically small Rossby numbers. It is found that as scales are decreased to reach the Zeman scale, anisotropy decreases (see \S \ref{s:aniso2}), and that isotropy is indeed recovered at that scale. {Note that this result, together with the previous results indicating that anisotropy increases with wavenumbers when the Zeman scale is smaller than the dissipation scale, would indicate anisotropy in rotating flows is non-monotonic with wavenumber. If that is the case, this would not be} the first documented case of non-monotonic behavior in rotating flows. As an example, in \cite{bourouiba}, it was shown using numerical simulations that the degree of coupling between 2D and 3D modes varies non-monotonically with the Rossby number, especially in what these authors call the intermediate Rossby number regime, which is characterized by a positive transfer of energy from waves to vortices.

One can think of parameter regimes such that the Zeman scale lies in the inertial range and plays a relevant role in the dynamics, in particular for geophysical and astrophysical flows in which the Reynolds numbers are very large whereas Rossby numbers are small but not exceedingly so. The condition $\eta=l_{\Omega}$, where $\eta$ is the dissipation scale, leads to $\epsilon = \nu \Omega^2$, a parameter that involves the dimensionless combination $Ro^2 Re$. This parameter was already proposed by \cite{canuto} (see their Eq.~38 with the energy dissipation rate being expressed with the usual Kolmogorv law, $\epsilon\sim U_0^3/L_0$). In other words, small-scale isotropization should occur for flows with $Re \ge Ro^{-2}$, a common occurrence in geophysical and astrophysical flows unless one is dealing with rapid rotators. Finally note that  in our run, the ratio of the Zeman scale to the Kolmogorov dissipation scale is found to be $\approx 25$ (see below).

\section{Recovery of isotropy beyond the Zeman scale: numerical results} \label{s:res}

Before proceeding to give the results of the numerical simulation, let us first define more precisely the different energy spectra we shall be dealing with. Starting from the velocity auto-correlation function $U_{ij}({\bf k})$ written in Fourier space, with the assumption of homogeneity but not of isotropy, one can define several sub-spectra in order to take into account angular variations. In that spirit, we define the following based on the trace of the tensor, $U({\bf k})=U_{ii}({\bf k})$.
The axisymmetric energy spectrum in Cartesian coordinates is:
\begin{equation}
e(|{\bf k}_{\perp}|,k_{\parallel})=
    \sum_{\substack{
          k_{\perp}\le |{\bf k}\times \hat {\bf z}| < k_{\perp}+1 \\
          k_{\parallel}\le k_z < k_{\parallel}+1}} U({\bf k}) 
    = \int U({\bf k}) |{\bf k}| \sin \theta d \phi \  
    = e(|{\bf k}|, \theta) \ ,
\label{etheta} \end{equation}
with $\theta$ the co-latitude in Fourier space with respect to the vertical axis with unit vector $\hat {\bf z}$, and $\phi$ the longitude with respect to the $x$ axis. Similar definitions hold for $h(|{\bf k}_{\perp}|, \theta$), the axisymmetric helicity spectrum, which is based on the antisymmetric part of the velocity correlation tensor.

Note the axisymmetric spectrum can also be expressed in terms of polar coordinates $e(|{\bf k}|, \theta)$, with $k = |{\bf k}|$, and $k_{\perp}= |{\bf k}_\perp| = |{\bf k}| \sin \theta$; $k_{\parallel}$ refers to the component of ${\bf k}$ in the direction of the rotation axis. The first definition in Eq.~(\ref{etheta}) involving double sums, is the discrete expression used in the simulations, and corresponds to counting energy in all modes with the same values of $k_\parallel$ and $k_{\perp}$ (in shells of unit width) but different values of $\phi$; geometric factors in the integral are automatically considered in the sums by the definition of  $k_{\perp}$, and as the number of modes per shell in $k_{\perp}$ varies as the distance from the origin. The spectra given for different angular variations are obtained by plotting $e(|{\bf k}|, \theta)$ along a line making an angle $\theta$ with the vertical axis; $\theta=0$ corresponds to $e(|{\bf k}_{\perp}|=0,k_{\parallel})$, and $\theta=\pi/2$ to $e(|{\bf k}_{\perp}|,k_{\parallel}=0)$. Note that when plotting contour lines of axisymmetric spectra, a trigonometric factor $1/\sin \theta$ will be included so that circles obtain in the fully isotropic case. 

From the axisymmetric spectra above, one can define the so-called reduced spectra $E$ as a function of $k_{\perp}$, $k_{\parallel}$, and $k$ as:
\begin{equation}
E(k_{\perp}) = \int e(|{\bf k}_{\perp}|,k_{\parallel}) d k_{\parallel} \ , \ 
E(k_{\parallel})= \int e(|{\bf k}_{\perp}|,k_{\parallel}) d k_{\perp} \ , \ 
E(k) = \int e(|{\bf k}|, \theta) |{\bf k}| d \theta \  .
\label{reduced} \end{equation}
All the reduced spectra have the physical dimension of an energy density, i.e., summed over wavenumber they yield, through Parseval's theorem, the total energy $\langle |{\bf u}|^2\rangle /2$.

\begin{figure} \begin{center}
\includegraphics[width=13.0cm]{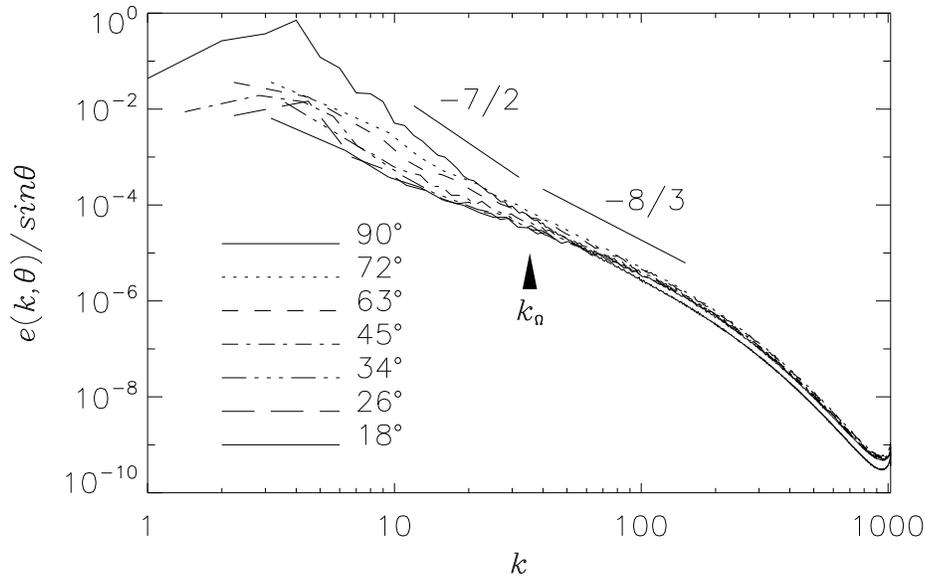}
\caption{Angular distribution of energy spectra for different co-latitudes $\theta$, averaged for $5\le t \le 6$; the Zeman wavenumber is indicated with an arrow. Slopes $-7/2$ (corresponding to $\sim k^{-5/2}$ scaling in units of the reduced energy spectra) and $-8/3$ (corresponding to Kolmogorov $\sim k^{-5/3}$ scaling) are shown as a reference. Note the recovery of isotropy beyond $k_{\Omega}$.}
\label{fig:ang}
\end{center} \end{figure}  

With these definitions, we now proceed to present the results of the numerical simulation. In Figure \ref{fig:spec} is shown the product of the reduced Fourier spectra $k_{\perp}^4 E(k_{\perp}) H(k_{\perp})/k_F$ compensated by $k_{\perp}^4$ and normalized by $k_F$, in terms of $k_{\perp}$. Unless explicitly stated, the spectra and fluxes presented in this and subsequent plots are averaged over the time interval $5\le t\le 6$, once stationarity is established in the small scales. Super-imposed on the graphs are two lines representing respectively compensated $k_{\perp}^{-4}$ and $k_{\perp}^{-10/3}$ power laws, and the arrow gives the location of the Zeman wavenumber $k_{\Omega}=2\pi/\ell_{\Omega}\approx 35$, computed from the imposed rotation and the energy flux. It is clear from Figure \ref{fig:spec} that we obtain an $e+h=4$ regime (see Eq.~\ref{zeman_G2b}) dominated by helicity transfer to small scales with a wave-induced transfer time at scales larger than $\ell_{\Omega}$, and that we recover dual Kolmogorov spectra for both energy and helicity at scales smaller than $\ell_{\Omega}$. Similar results obtain for the isotropic spectra. Also note the good agreement between the phenomenological evaluation of $\ell_{\Omega}$ and the numerical result. 

At these times, helicity is globally close to maximal, as measured for example on integrated quantities, e.g., $H/(k_F E)=0.88$ at $t=6$. The inertial range indices of energy and helicity are equal to $e=2.2$ and $h=1.8$ at scales larger than $\ell_{\Omega}$, with both close to $5/3$ at smaller scales where isotropy is recovered. However, it is worth noting that the relative helicity (measuring the relative alignment of velocity and vorticity) is 92\% at the injection scale, and 14\% at the Zeman scale when examining the isotropic spectra. A word of caution should be given here: for an anisotropic dynamical evolution, the isotropic spectra can be misleading since they integrate over spherical shells that are not equally populated (the $k_{\parallel}$ direction is less populated because of the quasi-bi-dimensionalization of the flow, at least up to $k_{\Omega}$). Yet one could ask why the helicity decreases (isotropically) so substantially, in relative terms, in the helically-dominated large-scale direct cascade. This may be related to the fact that, in such a flow, the helicity is concentrated in so-called Beltrami Core Vortices (BCVs), with $k_{\parallel}\approx 0$ \citep{1536b}. Such structures are quasi-two-dimensional fully helical tubes in the direction of the imposed rotation; they are concentrated at the scale of the forcing and are embedded in a tangle of thinner and more turbulent vortex filaments. With the velocity and the vorticity aligned in the BCVs, the columns are laminar and long-lived, evolving on a dissipative time-scale, whereas the rest of the flow may contain only marginal amounts of helicity on average. The BCVs could be viewed as Taylor columns, except that helicity plays an essential role in their structure. Wave propagation is observed along these structures, reminiscent of the linear wave propagation phenomenon invoked in \cite{davidson_2}. By construction, Beltrami vortices will survive, and moderately helical structures will transfer energy and helicity to small scale. Indeed, helicity is noticeably more intermittent than energy, as shown in \cite{1536b} by examining the probability distribution functions of velocity and helicity increments.

\begin{figure} \begin{center}
\includegraphics[width=13.0cm]{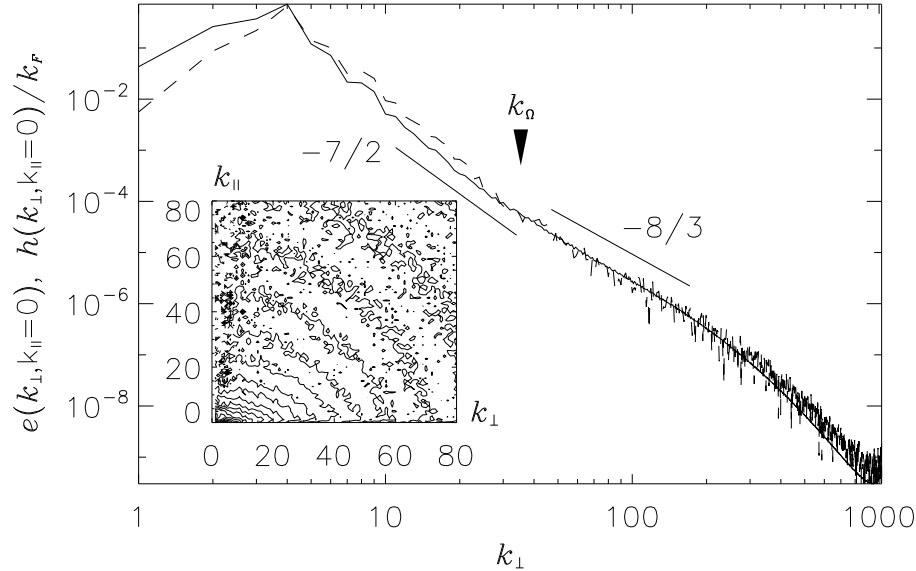}
\caption{Anisotropic spectra $e(k_{\perp}, k_{\parallel}=0)$ (solid line) and $h(k_{\perp}, k_{\parallel}=0)$ (dashed line, normalized by $k_F$) and averaged for $5\le t\le 6$; the slopes $-7/2$ and $-8/3$ are shown as references. Note the break-down of the helicity spectrum at the Zeman wavenumber (indicated by an arrow). The inset shows isolines of the axisymmetric helicity spectrum $h(k_{\perp}, k_{\parallel})/\sin \theta$ for small wavenumbers. Note the ellipsoidal isolines with lack of sign fluctuations for $|{\bf k}|<k_\Omega \approx 35$, and the more circular lines, with rapid fluctuations in sign, indicated by the increasingly noisier curves, for $|{\bf k}|>k_\Omega \approx 35$.}
\label{fig:para} \end{center} \end{figure}
  
In the inset of Figure \ref{fig:spec} is given the time evolution of a few individual modes (see caption), showing that the large scales are growing in time due to the inverse cascade of energy, whereas for the scales smaller than $2\pi/k_F$ but either larger or smaller than the Zeman scale $\ell_{\Omega}$, a quasi-steady state has already been reached at the times used for the analysis ($5 \le t \le 6$).

Note that, as a rough measure of large scale anisotropy,  one can compute the ratio of energy in the slow manifold (the 2D modes with $k_{\parallel}=0$) to the total energy, $E_{2D}/E$, as shown in Figure \ref{fig:new} (dashed line). In the run with $Ro\gg 1$, i.e., with negligible rotation, $E_{2D}/E\approx 0.25$ (also corresponding to $t=0$ of the run with rotation), while in the run with rotation $E_{2D}/E\approx 0.72$, i.e., almost three times larger and with most of the energy in 2D modes. In Figure \ref{fig:new} (solid line) is also shown the ratio of the perpendicular to the parallel integral scales of the flow. These scales are defined from the reduced spectra as
\begin{equation}
L_{\parallel} = 2\pi \frac{\int k_{\parallel}^{-1} E(k_{\parallel}) 
    dk_{\parallel}}{\int E(k_{\parallel}) dk_{\parallel}}  \ \ , \ \
    L_{\perp} = 2\pi \frac{\int k_{\perp}^{-1} E(k_{\perp}) 
    dk_{\perp}}{\int E(k_{\perp}) dk_{\perp}} .
\label{scales}\end{equation}
While at $t=0$ (i.e., in the turbulent state of the run with negligible rotation) $L_\perp/L_\parallel \approx 1$, corresponding to a flow which is isotropic, as time evolves this ratio fluctuates around 2, indicating that large scale anisotropy develops as a result of the increased rotation. Since in both cases ($t=0$ and $t>0$), the forcing is the same and since we started the strongly rotating run from the last state of the run with $Ro\gg 1$, we can safely conclude that the source of anisotropy displayed for example in Figure \ref{fig:new} is not due to the forcing but is due rather to the nonlinear dynamics of this strongly rotating and turbulent flow.

{A more complete picture of the development of anisotropy as the result of rotation is given by the axisymmetric energy spectrum $e(k_\perp,k_\parallel)$ at early and late times (see Figure \ref{fig:dev_aniso}). While at $t=0$ the contour levels of $e(k_\perp,k_\parallel)/\sin\theta$ are nearly circular, indicating isotropy at all wavenumbers, the spectrum at late times shows elliptical isocontours for small wavenumbers, with more energy in modes with $k_\parallel \approx 0$. As the wavenumber increases, the contour levels become less elliptical, an indication that small scales are more isotropic. This re-isotropisation at small scales will be studied in more detail below using the angular distribution of energy for different co-latitudes $\theta$ in Fourier space.}

It is perhaps remarkable that the fluxes of energy and helicity, given in Figure \ref{fig:flux}, do not show any transition at the Zeman scale: the fluxes continue to be constant in both direct cascades, with an overall excess of normalized helicity flux, as expected for this value of the Rossby number, an excess which persists in the isotropy range. There is a  smooth apparently seamless transition from one (anisotropic, Coriolis) subrange to the other (isotropic, Kolmogorov) subrange, indicative of a dual cascade with constant energy and helicity fluxes. The constancy of flux persists up to slightly beyond  wavenumber $k\approx 100$, after which dissipative effects are being felt (with the maximum wavenumber resolved in the computation $k_{max}=1024$). Note again that no hyperviscosity scheme is employed in this direct numerical simulation that would prolong the second isotropic range to small scales closer to the cut-off. Finally, at wavenumbers smaller than the forcing, an incipient inverse cascade of energy, indicated by the negative energy flux, can also be observed {even though it occurs over a narrow range of wavenumbers}.  

In the inset of Figure \ref{fig:flux} is given the relative helicity $r(k)=H(k)/[kE(k)]$. One observes that the relative helicity never reaches the maximum allowed value of unity, although it is quite close to it at the forcing wavenumber. At wavenumbers immediately  {larger} than $k_\Omega$ the relative helicity decreases with increasing wavenumber, approaching the well-known decrease $r(k)\sim 1/k$ of isotropic helical turbulence for wavenumbers close to and larger than $k_\Omega$, but smaller than $k\approx 100$. A clearer, but shallower, power law ($\sim k^{-1/2}$) is found in the dissipative range, beyond $k\approx 100$. This latter behavior has been reported before in simulations of isotropic helical turbulence \citep{mininni08}: it is known that point-wise helicity grows in a turbulent flow \citep{matthaeus}, and that 
{small-scale vortical} structures in non-rotating turbulence tend to be helical.

Figure \ref{fig:ang} presents the angular distribution of the energy spectrum $e(k_\perp,k_\parallel)$ defined in Eq. (\ref{etheta}) (see also  Figure \ref{fig:dev_aniso}): there is a wide distribution of power law indices from rather shallow for small co-latitudes to steep for large co-latitudes ($k_\parallel = 0$ modes). Again, at scales smaller than the Zeman scale, isotropy recovers with all angular spectra collapsing {to} the Kolmogorov spectrum (which corresponds to a $-8/3$ law in those units) at those scales. A detail of the so-called slow manifold spectra ($k_{\parallel}=0$) is provided in Figure \ref{fig:para} for the energy (solid line) and the helicity (dashed line). The terminology of ``slowness'' is of course linked to the fact that, for $k_{\parallel}=0$, the inertial wave frequency is zero. Note for the helicity the consistent spectrum up to $k_{\Omega}=2\pi/\ell_{\Omega}$ and the strong fluctuations in the helicity spectrum past that scale, with no definite sign (however, the fully integrated helicity spectrum $H(k)$ is positive all the way to the dissipation scale, not shown). The $-7/2$ power law indicated as a reference in the Coriolis subrange of both figures corresponds to the prediction for the energy spectrum that follows from the $e+h=4$ relation in the case of maximal helicity: the $-5/2$ slope in the isotropic energy spectrum corresponds to $-7/2$ in the axisymmetric spectra. In the inset of Figure \ref{fig:para}, contour levels of the axi-symmetric helicity spectrum are given, with a linear zoom on the smallest wavenumbers: for scales larger than the Zeman scale, anisotropy shows up in the form of elliptical contours, whereas past that scale, there is a breakdown of coherence of the angular helicity spectrum. 

Finally, we note that in all the spectra displayed in Figs. \ref{fig:ang} and \ref{fig:para}, no bottlenecks are visible: such a feature, i.e., the increase of energy at small scale just before the onset of the dissipation range, is prominent in many simulations of isotropic (non-rotating) turbulence (see, e.g., \citealt{ishihara} for a recent review of DNS of isotropic turbulence). The absence (or weakness) of the bottleneck here may be associated with non-local interactions between widely separated scales induced by the imposed rotation \citep{alex}.

\section{Conclusions}\label{s:conclu}

We have analyzed in this paper the recovery of isotropy in a turbulent rotating flow with helical forcing. The helical regime found previously in \cite{rot512,1536a} also appears at intermediate scales, followed by an isotropic Kolmogorov dual spectrum for energy and helicity at small scale. This indicates that the small-scale isotropic eddies do not destroy the phase-coherence of the helical columnar structures at large scale (through up-scaling interactions), but that their interactions {and structure}, as one goes to smaller scales, become predominant and isotropize, {with strong small-scale variations (not shown).} 

In shell models of turbulence as studied by \cite{plunian} to examine the return to isotropy at small scales in rotating flows in the presence of an imposed magnetic field and at low magnetic Prandtl number, it was shown that the shortest time-scale determined which spectrum develops. Here as well, we confirm the phenomenological analysis of \cite{zeman} that led to the prediction of a return to isotropy in the small scales when the eddy turn-over time becomes shorter than the inertial time beyond $\ell_{\Omega}$, including in the presence of helicity. 

 When computed at the Zeman scale, the micro-Rossby number is found to be of order unity, as mentioned earlier. Note that the Rossby number appears to be in the range of values given in \cite{bourouiba} for ``intermediate'' behavior in which the inverse cascade is strongest. This inverse cascade does indeed disappear, at least at lowest order, in the weak turbulence derivation of rotating flows performed by \cite{galtier_WT}, and yet it plays an essential role in leading to the dual helicity-energy cascade to small scales. Hence, it would be of interest, but costly, to perform a run similar to the one presented here at a smaller Rossby number, yet resolving better all direct and inverse dynamical ranges. In this context, large eddy simulations could possibly help in providing a model of rotating turbulent flows (see, e.g., \citealt{julien}).

We would expect that, in the case of a non-helical forcing such as the Taylor-Green flow, the same behavior would occur for the same range of parameters, with a $k^{-2}$ spectrum dependence for the energy at scales of the order of the forcing scale and smaller (and a $k_\perp^{-3}$ scaling of the modes with $k_\parallel=0$ in the axi-symmetric energy spectrum), before re-isotropization beyond the Zeman scale. The reason is that the expression for the Zeman scale does not depend on helicity and hence, presumably, the dynamics of the isotropic Kolmogorov subrange should not either, since it is known that in the non-rotating case turbulence dynamics with or without helicity is of the Kolmogorov type. However, there would be no discernible cascade of helicity in the Kolmogorov range, since it would likely be fluctuating around zero, contrary to what is observed here with maximally helical forcing.

Many additional questions on the statistical properties of the flow described in this paper need to be addressed, especially considering the role of anisotropy. {In particular, the simulation described here shows decreasing anisotropy with increasing wavenumber at all scales, while recent experiments of freely decaying rotating turbulence \citep{Lamriben} observed increased anisotropy at smaller scales (see also \citealt{bellet_06} for numerical and theoretical results). The reasons for this discrepacy are not clear, and {may} be related with differences between freely decaying and forced rotating flows (e.g., associated with the development of a quasi-steady inverse cascade of energy in the forced case), or with non-monotonicity of anisotropy with respect to wavenumber. If the case is the latter, it should be noted that as a result of limitations in resolution and computing power, our simulation has a short Coriolis subrange, and even larger scale separations may be needed to observe anisotropy increase with wavenumber before the Zeman scale is reached.}

{Finally, other questions to be addressed include properties of} small-scale statistics and intermittency. Indeed, several recent laboratory experiments of rotating turbulence \citep{simand,baroud03,moisy,van} as well as numerical simulations \citep{muller, rot512, 1536b} have found that rotating turbulence in the direct energy cascade is closer to self-similarity than homogeneous and isotropic turbulence, with the exponents of $p$-th order structure functions following $\zeta_p \approx \gamma p$. The exponent $\gamma$ is found to be either close to $1/2$ (compatible with a $k^{-2}$ non-helical spectrum), or in the range $0.7\le \gamma \le 0.75$ in the presence of helicity, compatible with an energy spectrum slightly shallower than $k^{-5/2}$.  Other issues arise as to the universality of the scaling presented here, for example when one varies the relative helicity of the forcing term. These questions are left for future work.

\begin{acknowledgments}
Computer time was provided by NSF under TeraGrid Grant No.~TG-PHY100029. NCAR is sponsored by the National Science Foundation. We acknowledge support from NSF-CMG Grant No.~1025188, and PDM acknowledges support from PICT Grant No. 2007-02211, UBACYT Grant No.~20020090200692, PIP Grant No. 11220090100825, and from the Carrera del Investigador Cient\'{\i}fico of CONICET.
\end{acknowledgments}

\bibliographystyle{jfm}

\begin{thebibliography}{23}

\bibitem[Archontis, Dorch \&  Nordlund(2003)]{Archontis03}
\textsc{Archontis, V., Dorch, S. B .F. \&  Nordlund, \AA.}
2003 Dynamo action in turbulent flows.
\emph{Astronom. Astrophys.} \textbf{410}, 759--766.

\bibitem[Baerenzung et al.(2011)]{julien}
\textsc{Baerenzung, J., Rosenberg, D., Mininni, P.D., \& Pouquet, A.}
2011 Helical Turbulence Prevails over Inertial Waves in Forced Rotating Flows at High Reynolds and Low Rossby Numbers
\emph{J. Atmos. Sci.} \textbf{68}, 2757--2770.

\bibitem[Baroud et al.(2003)]{baroud03}
\textsc{Baroud, C. N., Plapp, B. B., Swinney, H. L \& She, Z.-S.}
2003 Scaling in three-dimensional and quasi-two-dimensional rotating turbulent flows.
\emph{Phys. Fluids} \textbf{15}, 2091--2104.

\bibitem[Bartello, M\'etais \&  Lesieur(1994)]{bartello_CS}
\textsc{Bartello, P., M\'etais, O. \&  Lesieur, M.} 
1994 Coherent structures in rotating three-dimensional turbulence.
\emph{J. Fluid Mech.} \textbf{273}, 1--29.

\bibitem[Bellet et al.(2006)]{bellet_06}
\textsc{Bellet, F., Godeferd, F.S., Scott, F.S. \& Cambon, C.} 
2006 Wave turbulence in rapidly rotating flows.
\emph{J. Fluid Mech.} \textbf{562}, 83--121.

\bibitem[van Bokhoven et al.(2009)]{van}
\textsc{van Bokhoven, L.J., Clercx, H.J.H., van Heijst, G.J.F., \& Trieling, R.R.}
2009  Experiments on rapidly rotating turbulent flows.
\emph{Phys. Fluids}
\textbf{21}, 096601, 20 pages.

\bibitem[Bourouiba \& Bartello(2007)]{bourouiba}
\textsc{Bourouiba, L., \& Bartello, P.}
2007 The intermediate Rossby number range and two-dimensional-three-dimensional transfers in rotating decaying homogeneous turbulence.
\emph{J. Fluid Mech.} \textbf{587}, 139--161.

\bibitem[Brissaud et al.(1973)]{brissaud}
 \textsc{Brissaud, A., Frisch, U., L\'eorat, J.. Lesieur, M., \& Mazure, A.}
1973 Helicity cascades in fully developed isotropic turbulence.
 \emph{J. Fluid Mech.} \textbf{202}, 295--317.

\bibitem[Cambon \& Jacquin(1989)]{cambon_89}
\textsc{Cambon, C. \& Jacquin, L.}
1989 Spectral approach to non isotropic turbulence subjected to rotation.
\emph{J. Fluid Mech.} \textbf{202}, 295--317.

\bibitem[Cambon, Mansour, \& Godeferd(1997)]{cambon97}
\textsc{Cambon, C., Mansour, N.~N., \& Godeferd, F.~S.}
1997 Energy transfer in rotating turbulence.
\emph {J. Fluid Mech.} \textbf{337}, 303--332.

\bibitem[Cambon \& Scott(1999)]{cambon_rev} 
\textsc{Cambon, C., \& Scott, J.F.}
1999 Linear and Non-Linear Models of Anisotropic Turbulence.
\emph{Ann. Rev. Fluid Mech.} \textbf{11}, 1--53.

\bibitem[Cambon, Rubinstein \& Godeferd(2004)]{cambon_04}
\textsc{Cambon, C., Rubinstein, R. \& Godeferd, F.S.}
2004 Advances in wave turbulence: rapidly rotating flows.
\emph{New J. Phys.} \textbf{6}, 73, 29 pages.

\bibitem[Canuto \& Dubovikov(1997)]{canuto}
\textsc{Canuto, V.M. \& Dubovikov, M.S.}
1997 A dynamical model for turbulence. V. The effect of rotation.
\emph{Phys. Fluids} \textbf{9}, 2132--2140.

\bibitem[Chakraborty(2007)]{chakra07b}
\textsc{Chakraborty, S.}
2007 Signatures of two-dimensionalization of 3D turbulence in the presence of rotation.
\emph{European Physical Letters} \textbf{79}, 14002, 5 pages.

\bibitem[Chakraborty \& Batthacharjee(2007)]{chakra07}
\textsc{Chakraborty, S. \& Batthacharjee, J.K.}
2007 Third-order structure function for rotating in three-dimensional homogeneous turbulent flow.
\emph{Phys. Rev. E} \textbf{76}, 036304, 6 pages.

\bibitem[Chen, Chen \& Eyink(2003)]{chen}
\textsc{Chen, Q., Chen, S., \& Eyink, G.}
2003 The joint cascade of energy and helicity in three-dimensional turbulence.
\emph{Phys. Fluids} \textbf{15}, 361--374.

\bibitem[Chen et al.(2005)]{chen05}
\textsc{Chen, Q., Chen, S.,  Eyink, G.L. \& Holm, D.D.}
2005 Resonant interactions in rotating homogeneous three-dimensional turbulence.
\emph{J. Fluid Mech.} \textbf{542}, 139--164.

\bibitem[Davidson, Staplehurst \& Dalziel(2006)]{davidson_2}
\textsc{Davidson, P.A., Staplehurst, P.J., \& Dalziel, S.B.}
2006 On the evolution of eddies in a rapidly rotating system.
\emph{J.Fluid Mech.} \textbf{557}, 135--144.

\bibitem[Dubrulle  \& Valdetarro(1992)]{dubrulle}
\textsc{Dubrulle, B., \& Valdetarro, L.}
1992 Consequences of rotation in energetics of accretion disks.
\emph{Astronom. Astrophys.} \textbf{263}, 387--400.

\bibitem[Galtier(2003)]{galtier_WT} 
\textsc{Galtier, S.}
2003 Weak Inertial-Wave Turbulence Theory.
\emph{Phys. Rev. E} \textbf{68}, 015301 (R), 4 pages.

\bibitem[G\'omez, Mininni \& Dmitruk(2005)]{Gomez05a}
\textsc{G\'omez, D.O, Mininni, P.D., \& Dmitruk, P.}
2005 Parallel simulations in turbulent MHD.
\emph{Phys. Scripta} \textbf{T116}, 123--127.

\bibitem[Hopfinger, Browand \& Gagne(1982)]{hopfinger82}
\textsc{Hopfinger, E. J. Browand, F. K.  \& Gagne, Y.  }  
1982 Turbulence and waves in a rotating tank. 
\emph{J. Fluid Mech.} \textbf{125}, 505--534.

\bibitem[Hopfinger \& van Heijst(1993)]{hopfinger93} \textsc{Hopfinger, E. J. \& van Heijst, G. J.}
1993 Vortices in Rotating Fluids. 
\emph{Ann. Rev. Fluid Mech.} \textbf{25}, 241--289.

\bibitem[Ishihara,  Gotoh \& Kaneda(2009)]{ishihara}
\textsc{Ishihara,  T., Gotoh, T., \& Kaneda, Y.}
2009 Study of High-Reynolds Number Isotropic Turbulence by Direct Numerical Simulation.
\emph{Ann. Rev. Fluid Mech.}
\textbf{41}, 165-180.

\bibitem[Koprov et al.(2005)]{koprov}
\textsc{Koprov, B.M., Koprov, V. M., Ponomarev, V.M., \& Chkhetiani, O.G.}
2005 Experimental Studies of Turbulent Helicity and Its Spectrum in the Atmospheric Boundary Layer.
\emph{Doklady Physics}
\textbf{50}, 419--422. Translated from \emph{Doklady Akademii Nauk} {\bf 403}, 627--630 (2005).

\bibitem[Kraichnan(1973)]{rhk}
\textsc{Kraichnan, R.H.}
1973 Helical turbulence and absolute equilibrium.
\emph{J. Fluid Mech.} \textbf{59}, 745--752.

\bibitem[Lamriben, Cortet \& Moisy (2011)]{Lamriben}
\textsc{Lamriben, C., Cortet, P.-P., \& Moisy, F.}
2011 Direct Measurements of Anisotropic Energy Transfers in a Rotating Turbulent Experiment.
\emph{Phys. Rev. Lett.} \textbf{107}, 024503, 4 pages.

\bibitem[Lilly(1986)]{lilly}
\textsc{Lilly, D.}
1986 The Structure, Energetics and Propagation of Rotating Convective Storms. Part II: Helicity and Storm Stabilization.
\emph{J. Atmos. Sci.} \textbf{43}, 126--140.

\bibitem[Matthaeus et al.(2008)]{matthaeus}
\textsc{Matthaeus, W.H., Pouquet, A., Mininni, P.D., Dmitruk, P., \& Breech, B.}
2008 Rapid directional alignment of velocity and magnetic field in 
magnetohydrodynamic turbulence.
\emph{Phys. Rev. Lett.} \textbf{100}, 085003, 4 pages.

\bibitem[Mininni, Alexakis, \& Pouquet(2008)]{mininni08}
\textsc{Mininni, P.D., Alexakis, A., \& Pouquet, A.}
2008 Nonlocal interactions in hydrodynamic turbulence at high Reynolds numbers: The slow emergence of scaling laws.
\emph{Phys. Rev. E} \textbf{77}, 036306, 9 pages.

\bibitem[Mininni \& Pouquet(2009)]{rot512}
\textsc{Mininni, P.D. \& Pouquet, A.}
2009 Helicity cascades in rotating turbulence.
\emph{Phys. Rev. E} \textbf{79}, 026304, 7 pages.

\bibitem[Mininni, Alexakis \& Pouquet(2009)]{alex}
\textsc{Mininni, P.D., Alexakis, A. \& Pouquet, A.}
2009 Scale interactions and scaling laws in rotating flows at moderate Rossby numbers and large Reynolds numbers.
\emph{Phys. Fluids.} \textbf{21}, 015108, 14 pages.

\bibitem[Mininni \& Pouquet(2010a)]{1536a}
\textsc{Mininni, P.D. \& Pouquet, A.}
2010a Rotating helical turbulence. Part I. Global evolution and spectral behavior.
\emph{Phys. Fluids} \textbf{22}, 035105, 9 pages.
  
\bibitem[Mininni \& Pouquet(2010b)]{1536b}
\textsc{Mininni, P.D. \& Pouquet, A.}
2010b Rotating helical turbulence. Part II. Intermittency, scale invariance and structures.
\emph{Phys. Fluids} \textbf{22}, 035106, 10 pages.

\bibitem[Mininni et al.(2011)]{ghybrid}
\textsc{Mininni, P.D. Rosenberg, D., Reddy, R., \& Pouquet, A.}
A hybrid MPI-OpenMP scheme for scalable parallel pseudospectral computations for
fluid turbulence.
 \emph {Parallel Computing}, {\textbf 37}, 316--326 (2011). 

\bibitem[Moffatt \& Tsinober(1992)]{moffatt_tsinober}
\textsc{Moffatt, H.K. \& Tsinober, A.}
1992 Helicity in laminar and turbulent flow.
\emph{Ann. Rev. Fluid Mech.} \textbf{24}, 281--312.

\bibitem[Moisy et al.(2011)]{moisy11}
\textsc{Moisy, F., Morize, C., Rabaud, M. \& Sommeria, J.}
2011 Decay laws, anisotropy and cyclone-anticyclone asymmetry in decaying rotating turbulence.
\emph{J. Fluid Mech.} \textbf{666}, 5--35.

\bibitem[M\"uller \& Thiele(2007)]{muller}
\textsc{M\"uller, W.-C. \& Thiele, M.}
2007 Scaling and energy transfer in rotating turbulence.
\emph{Europhys. Lett.} \textbf{77}, 34003, 5 pages.

\bibitem[Nazarenko \& Schekochihin(2011)]{naza}
\textsc{Nazarenko, S.V.. \& Schekochihin, A.A.}
2011 Critical balance in magnetohydrodynamic, rotating and stratified turbulence: towards a universal scaling conjecture. 
\emph{J. Fluid Mech.} \textbf{677}, 134--153.

\bibitem[Newell(1969)]{newell}
\textsc{Newell, A.C.}
1969 Rossby wave packet interactions.
\emph{J. Fluid Mech.} \textbf{35}, 255--271.

\bibitem[Orlandi \& Fatica(1997)]{orlandi}
\textsc{Orlandi, P. \& Fatica, M.}
1997 Direct simulations of turbulent flow in a pipe rotating about its axis.
\emph{J. Fluid Mech.} \textbf{343}, 43--72.

\bibitem[Plunian \& Stepanov(2010)]{plunian}
\textsc{Plunian, F. \& Stepanov, R.}
2010 Cascades and dissipation ratio in rotating magnetohydrodynamic turbulence
at low magnetic Prandtl number.
\emph{Phys. Rev. E} \textbf{82}, 046311, 6 pages.

\bibitem[Podvigina \& Pouquet(1994)]{podvigina}
\textsc{Podvigina O. \& Pouquet A.}
1994 On the Nonlinear Stability of the 1:1:1 ABC Flow.
{\em Physica D\/} {\bf 75}, 475--508.

\bibitem[Pouquet \& Patterson(1978)]{pouquet_patterson}
\textsc{Pouquet, A. \& Patterson, G.S.}
1978 Numerical simulation of helical magnetohydrodynamic turbulence.
\emph{J. Fluid Mech.} \textbf{85}, 305--323.

\bibitem[Pouquet \& Mininni(2010)]{phil_trans}
\textsc{Pouquet, A. \& Mininni, P.D.}
2010 The interplay between helicity and rotation in turbulence: implications for scaling laws and small-scale dynamics.
\emph{Phil. Trans. Roy. Soc.} \textbf{368}, 1635--1662.

\bibitem[Proudman(1916)]{proudman}
\textsc{Proudman, J.}
1916 On the motion of solids in a liquid possessing vorticity.
\emph{Proc. R. Soc. Lond. A} \textbf{92}, 408--424.

\bibitem[Seiwert, Morize \& Moisy(2008)]{moisy}
\textsc{Seiwert, J., Morize, C. \& Moisy, F.}
2008 On the decrease of intermittency in decaying rotating turbulence.
\emph{Phys. Fluids} \textbf{20}, 071702, 4 pages.

\bibitem[Simand(2002)]{simand}
\textsc{Simand, C.} 2002 \'Etude de la turbulence au voisinage d'un vortex. Th\`ese de Doctorat de l'Universit\'e de Lyon, \'Ecole Normale Sup\'erieure de Lyon.

\bibitem[Smith, Chasnov \& Waleffe(1996)]{smith96}
\textsc{Smith, L., Chasnov, R. \& Waleffe, F.}
1996 Crossover from Two- to Three-Dimensional Turbulence. 
\emph{Phys. Rev. Lett.} \textbf{77}, 2467--2470.

\bibitem[Smith \& Waleffe(1999)]{smith99}
\textsc{Smith, L., \& Waleffe, F.}
1999 Transfer of energy to two-dimensional large scales in forced, rotating three-dimensional turbulence.
\emph{Phys. Fluids} \textbf{11}, 1608--1622.

\bibitem[Taylor(1917)]{taylor}
\textsc{Taylor, G.I.}
1917 Motion of solids in fluids when the flow is not irrotational.
\emph{Proc. R. Soc. Lond. A} \textbf{93}, 92--113. 

\bibitem[Teitelbaum \& Mininni(2011)]{teitelbaum}
\textsc{Teitelbaum, T. \& Mininni, P.D.}
2011 Effect of helicity and rotation on the free decay of turbulent flows.
\emph{Phys. Rev. Lett.} \textbf{103}, 014501, 4 pages.

\bibitem[Waleffe(1993)]{waleffe}
\textsc{Waleffe, F.}
1993  Inertial transfers in the helical decomposition.
\emph{Phys. Fluids} \textbf{A5}, 677--685.

\bibitem[Zeman(1994)]{zeman}
\textsc{Zeman, O.}
1994 A note on the spectra and decay of rotating homogeneous turbulence.
\emph{Phys. Fluids} \textbf{6}, 3221--3223.

\bibitem[Zhou(1995)]{zhou}
\textsc{Zhou, Y.}
1995 A Phenomenological Treatment of Rotating Turbulence.
\emph{Phys. Fluids} \textbf{7}, 2092--2094.

\end{thebibliography}

\end{document}